\documentclass[aps,prc,twocolumn,showpacs]{revtex4}
\usepackage{ulem}
\usepackage{color}
\usepackage{graphicx}
\usepackage{epsfig}
\usepackage{subfig}
\usepackage{bm}
\usepackage{tabularx}
\usepackage{amssymb}
\usepackage[hidelinks,colorlinks=true,linkcolor=blue,citecolor=blue]{hyperref}
\def\be {\begin{equation}}
\def\ee {\end{equation}}
\def\nn {\nonumber}
\def\bea {\begin{eqnarray}}
\def\eea {\end{eqnarray}}

\def\vec#1{\mathchoice
	{\mbox{\boldmath $#1$}}
	{\mbox{\boldmath $#1$}}
	{\mbox{\boldmath $\scriptstyle #1$}}
	{\mbox{\boldmath $\scriptscriptstyle #1$}}
}

\newcommand{\om}{\omega}  
\newcommand{\vk}{\vec k}

\newcommand{\del}{\partial}

\begin{document}
	
	
	\title{
		Quantum expression for the electrical conductivity of massless quark matter and of the hadron resonance gas in the presence of a magnetic field
	}
	
	\author{Jayanta Dey$^{1}$\footnote{Presently at Indian Institute of Technology Indore}, Subhasis Samanta$^{2}$, Sabyasachi Ghosh$^{3}$  Sarthak Satapathy$^{4}$, }

	%
	\affiliation{$^1$ National Institute of Science Education and Research (NISER)
		Jatni 752050, India}
	\affiliation{$^{2}$Institute of Physics, Jan Kochanowski University, 25-406 Kielce, Poland}
	\affiliation{$^{3}$Indian Institute of Technology Bhilai, GEC Campus, Sejbahar, Raipur 492015, Chhattisgarh, India}

	\affiliation{$^4$ Dinabandhu Mahavidyalay, Bongaon, Dist: North 24 Parganas, Pin: 743235,
		West Bengal, India}
	%
	
	\begin{abstract}
		We have studied the classical and quantum expressions of electrical conductivity and their numerical estimation in the presence of a magnetic field for hadron resonance gas (HRG) and massless quark matter. Classical results of transport coefficients of HRG matter in the presence of a magnetic field were studied previously in {\it Phys.Rev.D 102 (2020) 1, 016016} using the standard relaxation time approximation in the Boltzmann equation. The estimations of transport coefficients for isotropic to anisotropic, massless quark matter to HRG matter transitions with the introduction of a magnetic field were made in the same reference. This led to an upper limit or Stefan-Boltzmann (SB)-type limit to non-perturbative domain transition of transport coefficients. In a similar context, the present work has concentrated on the classical to quantum transition of HRG transportation from high temperature and low magnetic field to low temperature and high magnetic field domain transition. 
		We have also compared the quantum modification of HRG results with that of massless quark matter, where we observed an opposite trend. A similar kind of quantum effect is also noticed between mesons and baryons due to their different particle distribution functions. Despite the fact that HRG contains both mesons and baryons, Landau quantization of its net magneto-thermodynamic phase-space reveals meson or boson-dominated quantum modification. That is why the quantum modification of HRG results reveals the opposite trend from the massless quark matter, which faces fermionic quantum modification. 

	\end{abstract}
	
	\maketitle
	\section{Introduction}
	In the presence of a magnetic field, quark gluon plasma (QGP) exhibits a number of interesting phenomena~\cite{Rafelski:1975rf,Voskresensky:1980nk,Schramm:1991ex,Schramm:1991vu,Kharzeev:2007jp,Kharzeev:2013jha}. The magnitude of magnetic field produced at RHIC for Au-Au collisions at $\sqrt{s} = 200 $ GeV is of the 
	order of $10^{19}$ Gauss and $eB\sim 10^{20}$ Gauss for Pb-Pb collisions at LHC. 
	This is much larger than the 
	$\Lambda_{QCD}^2 (\approx 2 \times 10^{18}$Gauss), where $\Lambda_{QCD} (\approx 0.2$ GeV) is the strong interaction scale. 
	The magnitude of the produced field is also very high compared to that of the neutron stars and magnetars, which is of the order of $10^{14}-10^{15}$ Gauss~\cite{Seiradakis:2004ew}.
	
	Understanding of the impact of this high magnetic field on different transport coefficients of QGP has recently appeared to be an important research topic within the community of heavy ion physics. 
	Recent Refs.~\cite{Li:2017tgi,Nam:2013fpa,Tawfik:2016ihn,Tuchin:2011jw,Ghosh:2018cxb,Dey:2019axu,Dey:2019vkn,Dash:2020vxk,Das:2019pqd,Chen:2019usj,Denicol:2018rbw,Mohanty:2018eja,Hattori:2017qih,Huang:2009ue,Agasian:2013wta,Agasian:2011st,Kurian:2018dbn,Hattori:2016cnt,Das:2019ppb,Kurian:2018qwb,Kurian:2017yxj,Harutyunyan:2016rxm,Kerbikov:2014ofa,Nam:2012sg,Rath:2019vvi,Das:2019wjg,Ghosh:2019ubc,Thakur:2019bnf,Chatterjee:2019nld,Kurian:2020qjr} 
	have gone through the calculations of different transport coefficients like 
	shear viscosity ~\cite{Li:2017tgi,Nam:2013fpa,Tawfik:2016ihn,Tuchin:2011jw,Ghosh:2018cxb,Dey:2019axu,Dey:2019vkn,Dash:2020vxk,Das:2019pqd,Chen:2019usj,Denicol:2018rbw}, 
	bulk viscosity ~\cite{Hattori:2017qih, Huang:2009ue, Agasian:2013wta, Agasian:2011st,Kurian:2018dbn}
	and electrical conductivity~\cite{Dey:2019axu,Dey:2019vkn,Dash:2020vxk,Das:2019pqd,Hattori:2016cnt,Das:2019ppb,Kurian:2018qwb,
		Kurian:2017yxj,Harutyunyan:2016rxm,Kerbikov:2014ofa,Nam:2012sg,Rath:2019vvi,Das:2019wjg,Ghosh:2019ubc,Thakur:2019bnf,Chatterjee:2019nld,Kurian:2020qjr} 
	in the presence of a magnetic field. Among them, electrical conductivity ($\sigma$) plays an important role in the lifetime of the magnetic field produced in heavy ion collisions (HIC).
	The produced magnetic field can exist for a longer time when the electrical conductivity  of RHIC or LHC matter is high~\cite{Tuchin:2015oka}.

	From the microscopic calculations, studied in earlier 
	Refs.~\cite{Dey:2019axu,Dey:2019vkn,Dash:2020vxk,Das:2019pqd,Hattori:2016cnt,Das:2019ppb,Kurian:2018qwb,Kurian:2017yxj,
		Harutyunyan:2016rxm,Kerbikov:2014ofa,Nam:2012sg,Rath:2019vvi,Das:2019wjg,Ghosh:2019ubc,Thakur:2019bnf,Chatterjee:2019nld,Kurian:2020qjr}, we can get temperature
	and magnetic field dependent values of the electrical conductivity of RHIC or LHC matter.
	If we analyze those investigations minutely, we can find mainly two classes of calculations: those done in	classical and quantum pictures.
	Among the Refs.~\cite{Dey:2019axu,Dey:2019vkn,Dash:2020vxk,Das:2019pqd,Hattori:2016cnt,Das:2019ppb,Kurian:2018qwb,Kurian:2017yxj,
		Harutyunyan:2016rxm,Nam:2012sg,Rath:2019vvi,Das:2019wjg,Ghosh:2019ubc,Thakur:2019bnf,Chatterjee:2019nld,Kurian:2020qjr},
	Refs.~\cite{Dey:2019axu,Dey:2019vkn,Dash:2020vxk,Das:2019pqd,Das:2019ppb,Harutyunyan:2016rxm,Das:2019wjg,Nam:2012sg,Thakur:2019bnf,Chatterjee:2019nld} have examined
	the classical expressions of electrical conductivity, whose multi-component values show an anisotropy in co-ordinate space. They have not, however, taken into account
	the quantum aspects of Landau quantization.
	In the context of neutron stars, Ref.~\cite{Harutyunyan:2016rxm} has used a similar classical expression of conductivity for magnetic fields of the order of $10^{14}$ Gauss, beyond which it is important to take into account the quantum effects, as done in Ref.~\cite{Potekhin:1999ur}. In the quantum domain, due to the Landau quantization, the motion of the charged particles gets quantized in the plane perpendicular to the magnetic field.
	Refs.~\cite{Hattori:2016cnt,Kurian:2018qwb,Kurian:2017yxj,Rath:2019vvi,Ghosh:2019ubc,Kurian:2020qjr} have
	considered this Landau quantization, where most of them~\cite{Hattori:2016cnt,Kurian:2018qwb,Kurian:2017yxj,Rath:2019vvi} have
	gone through the lowest Landau level (LLL) approximation which is applicable for the strong field limit.
	The LLL approximation reduces the multi-component structure of conductivity to a single component which is parallel to the magnetic field's direction. This component is referred to as longitudinal conductivity. The classical framework will be useful in the weak field range, while the LLL approximated quantum framework will be useful in the strong field limit.
	How does the multi-component classical expression of conductivity change during the weak to strong field transition?
	This question is attempted to be answered in the current work using the numerical benchmark of the ideal hadron resonance gas (HRG) model, which is well established as an alternative description of QCD thermodynamics within the hadronic temperature range~\cite{Braun-Munzinger:2003pwq}.
	
	Using the HRG model, Refs.~\cite{Dash:2020vxk,Das:2019wjg,Das:2019pqd} estimated the classical multi-component transport coefficients in the presence of a magnetic field.
	However, their quantum extensions have not yet been completed.
	The current work will fill the gap by focusing solely on electrical conductivity, because the mathematical extension of other transport coefficients is analogous. 
	
	The article is organized as follows: The next section (\ref{sec:Form}) addresses all working formulas of multi-components of conductivity at a finite magnetic field with a quick description of kinetic theory formalism based on the standared relaxation time approximation (RTA). This formalism section is divided into three subsections, where subsection (\ref{sec:el_B}) is devoted to the expressions of anisotropic conductivity components, which we call classical expressions, then in subsection (\ref{sec:QM}), their quantum extensions are addressed; and after that, in subsection (\ref{sec:HRG}), HRG version of classical and quantum expressions both are addressed, which behave as working formulae of the result section i.e. Sec.~(\ref{sec:Res}). In the result section, the classical to quantum transitions for the HRG model are graphically sketched and discussed. At the end, the investigation is summarised in Sec.~\ref{sec:Sum}.

	\section{Formalism}
	\label{sec:Form}
	Here, we will go over the formalism section step by step so that we can see how the expressions change from classical to quantum and then to the HRG model.
	We know that in the presence of a magnetic field, the conductivity tensors lose their isotropic nature, and we get different neumarical values or expressions for the parallel, perpendicular, and another component called the Hall conductivity. The anisotropic conductivity tensor can obtained using the relaxation time approximation (RTA) based kinetic theory  approach~\cite{Landau,Harutyunyan:2016rxm,Dey:2019axu}, whose expressions are considered here as classical (CL) expressions where the Landau quantization is not taken into account.  We are, however, considering quantum aspects of statistical mechanics by employing Fermi-Dirac (FD) and Bose-Einstein (BE) distribution functions for baryons and mesons, respectively. So, from the same perspective, the notation CL does not exactly mean classical expression. The derivation of these CL expressions is given in the following subsection (\ref{sec:el_B}) and the quantum (QM) expressions, where the Landau quantization is considered, are addressed in the subsequent subsection (\ref{sec:QM}). Finally, in the subsection (\ref{sec:HRG}), the HRG version of CL and QM expressions are formulated. 
	
	\subsection{Classical expressions of electrical conductivity} 
	\label{sec:el_B}
	For a detailed derivation of classical expressions (CL) of anisotropic conductivity reader can go through Refs.~\cite{Dey:2019axu,Dey:2019vkn,Harutyunyan:2016rxm,Landau}. However, for the sake of completeness, let us quickly go through the steps to the final expressions.
	
	Let us consider an electric field ${\vec E}=E_x {\hat x}$ is applied to a relativistic charged fermion/boson fluid, 
	for which a current density is obtained along the same direction ${\vec J}=J_x {\hat x}$.
	Hence, macroscopic Ohm's law can be written as
	\be
	J_x=\sigma_{xx}E_x~,
	\label{macro_E}
	\ee
	where $\sigma_{xx}$ is the electrical conductivity. In microscopic theory of dissipation, equilibrium distribution function of fermion/boson,
	\be
	f_0=\frac{1}{e^{\beta\om}\mp 1}~,
	\ee
	undergoes a small deviation ($\delta f$) driven by the electric field (neglecting other dissipative forces).
	So, the ansatz of $\delta f$ can be written as	
	\bea
	\delta f &\propto& \Big(\frac{\del f_0}{\del \om}\Big)
	\nn\\
	\delta f &=& -\vec{\alpha} \cdot \vec{E} ~\Big(\frac{\del f_0}{\del \om}\Big)
	\nn\\
	&=& (\alpha_x E_x)\beta f_0(1\mp f_0)~,
	\label{d_f}
	\eea
	where $\vec \alpha$	is an unknown coefficient that can be find usign the Boltzmann transport equation. Therefore, one can express (dissipative) current density as~\cite{Dey:2019axu,Dey:2019vkn,Landau,Harutyunyan:2016rxm} 
	\bea
	J_x &=&g{\tilde e}\int \frac{d^3\vk}{(2\pi)^3}\frac{k_x}{\om}\delta f
	\nn\\
	&=&\Big[g{\tilde e}\beta\int \frac{d^3\vk}{(2\pi)^3}\frac{k_x}{\om}\alpha_x f_0(1\mp f_0)\Big ]E_x~,
	\label{micro_E}
	\eea
	where $g$ is the degeneracy factor (excluding charge-flavor degeneracy), 
	${\tilde e}$ is electric charge and $\om=\{\vk^2+m^2\}^{1/2}$ is energy of the particle. 
	To find out $\alpha_x$, we take help of relaxation
	time approximated- relativistic Boltzmann equation (RTA-RBE),
	\bea
	-{\tilde e}{\vec E}\cdotp{\vec \nabla_k} f_0&=&-\delta f/\tau_c
	\nn\\
	\Rightarrow \delta f &=& \tau_c{\tilde e}{\vec E}\cdotp\frac{\vec k}{\om}
	\Big[\frac{\del f_0}{\del \om}\Big]
	\nn\\
	&=&\tau_c{\tilde e}E_x\Big(\frac{k_x}{\om} \Big)[\beta f_0(1\mp f_0)]~,
	\label{del_f}
	\eea
	where $\tau_c$ is called the relaxation time. Now, comparing Eq.~(\ref{del_f}) and (\ref{d_f}), we get
	\be
	\alpha_x=\tilde e \tau_c\frac{{k_x}}{\om}~.
	\ee
	Using the above expression of $\alpha_x$ in Eq.~(\ref{micro_E}) and comparing with Eq.~(\ref{macro_E}),
	we get expression of electrical conductivity which give rise to electric
	current in x direction as,
	\be
	\sigma_{xx}=g{\tilde e}^2 \beta\int \frac{d^3\vk}{(2\pi)^3}
	\tau_c\frac{k_x^2}{\om^2} f_0(1\mp f_0)~.
	\label{sig_B0}
	\ee

	Next, we will derive the electrical conductivity in the presence of a
	magnetic field ${\vec B}=B {\hat z}$,
	where the force term $\frac{d\vk}{dt}$ in RTA-RBE will be modified by the Lorentz force ($-{\tilde e} ({\vec E} +{\vec v}\times {\vec B})$) as 
	\bea
	{-\tilde e} ({\vec E} +\frac{\vec k}{\om}\times {\vec B})\cdot\nabla_k f_0 &=& \frac{-\delta f}{\tau_c}
	\nn\\
	{-\tilde e} ({\vec E} +\frac{\vec k}{\om}\times {\vec B})\cdot\Big(\frac{\vk}{\om}\Big)\frac{\del f_0}{\del\om} 
	&=& \frac{-\delta f}{\tau_c}~.
	\eea
	The second term on the left hand side gives null contribution  because of the vector identity $(\vk\times{\vec B})\cdot \vk={\vec B}\cdotp(\vk\times\vk)=0$, so we consider the contribution from the term $\nabla_k(\delta f)$ in RTA-RBE,
	\be
	{-\tilde e}{\vec E}\cdot\Big(\frac{\vec k}{\om}\Big)\frac{\partial f_0}{\partial \om} 
	- {\tilde e}(\frac{\vec k}{\om}\times {\vec B})\cdot \nabla_k(\delta f)
	=-\delta f/\tau_c~.
	\label{RBE_H_df}
	\ee
	Here, the dissipation is driven by $\vec E$ as well as $\vec B$. So, we consider $\delta f=-(\vec{k} \cdot \vec{F}) \frac{\del f_0}{\del \om}$, $\vec F$ is a function of $\vec E$ and $\vec B$.
	Now, using the standard vector identity,
	\bea
	\left(\frac{\vec k}{\om}\times {\vec B}\right)\cdot\nabla_k(\delta f)
	&=&-\left(\frac{\vec k}{\om}\times {\vec B}\right)\cdot \nabla_k(\vk\cdot {\vec F})~ \frac{\partial f_0}{\partial \om}
	\nn\\
	&=&-\left(\frac{\vec k}{\om}\times {\vec B}\right)\cdot{\vec F}~ \frac{\partial f_0}{\partial \om}
	\nn\\
	&=&-\frac{\vec k}{\om}\cdot ({\vec B}\times{\vec F}) ~\frac{\partial f_0}{\partial \om}~,
	\label{del_f_expand}
	\eea
	in Eq.~(\ref{RBE_H_df}), we get
	\be
	\Big(\frac{\vec k}{\om}\Big)\cdot\Big[-{\tilde e}{\vec E} + {\tilde e}({\vec B}\times{\vec F})\Big]
	=\vk\cdot {\vec F}/\tau_c~.
	\label{RTA_RBE_B}
	\ee
	In general, we can consider
	\be
	{\vec F}= (A_x {\hat x} + A_z{\hat z} + A_y({\hat x}\times{\hat z}) ), 
	\ee
	for which Eq.~(\ref{RTA_RBE_B}) becomes
	\begin{widetext}
		\be
		\frac{\tau_c}{\om}\Big[-{\tilde e}E_x{\hat x} + {\tilde e}B{\hat z}\times\Big(A_x {\hat x} + A_z{\hat z} 
		- A_y{\hat y}\Big)\Big]
		=\Big(A_x {\hat x} + A_z {\hat z} - A_y {\hat y}\Big)
		%
		\label{Bolz_phi_B}
		\ee
	\end{widetext}
	%
	%
	Equating the coefficients of ${\hat x}$, ${\hat z}$ and ${\hat y}$ of Eq.~(\ref{Bolz_phi_B}), 
	we get
	\bea
	A_z &=& 0
	\nn\\
	A_x &=&-\tilde{e} \frac{1}{1+(\tau_c/\tau_B)^2}\frac{\tau_c}{\om}E_x
	\nn\\
	A_y &=& \tilde{e}\frac{\tau_c/\tau_B}{1+(\tau_c/\tau_B)^2}\frac{\tau_c}{\om}E_x
	\eea
	where $\tau_B=\om/({\tilde e}B)$ is the inverse of cyclotron frequency.
	So, final form of the deviation becomes
	\bea
	\delta f&=&-{\vec k}\cdot\Big\{-\frac{{\tilde e}\tau_c}{\om}\Big({\hat x}+\frac{\tau_c}{\tau_B}{\hat y}\Big)\Big\}
	\frac{1}{1+(\tau_c/\tau_B)^2}~\frac{\del f_0}{\del \om}
	\nn\\
	&=&-{\tilde e}\tau_c\Big(\frac{k_x}{\om}+\frac{k_y}{\om}\frac{\tau_c}{\tau_B}\Big)E_x
	\frac{1}{1+(\tau_c/\tau_B)^2}~\beta f_0(1-f_0)
	\nn\\
	\eea
	Now, using this $\delta f$ in matrix form of Ohm's law,
	\be
	\left(
	\begin{array}{c}
		J_x \\
		J_y 
	\end{array}
	\right) =
	\left(
	\begin{array}{cc}
		\sigma_{xx} & \sigma_{xy} \\
		\sigma_{yx} & \sigma_{yy}
	\end{array}
	\right)
	\left(
	\begin{array}{c}
		E_x \\
		0 
	\end{array}
	\right)
	\ee
	one can obtain 
	\bea
	\sigma_{xx}&=&g{\tilde e}^2 \beta\int \frac{d^3\vk}{(2\pi)^3}
	\tau_c\frac{1}{1+(\tau_c/\tau_B)^2}\frac{k_x^2}{\om^2} f_0(1\mp f_0)
	\nn\\
	&=&g{\tilde e}^2 \beta\int \frac{d^3\vk}{(2\pi)^3}
	\tau_c\frac{1}{1+(\tau_c/\tau_B)^2}\frac{\vk^2}{3\om^2} f_0(1\mp f_0)
	\label{sig_xxCl}
	\\
	\sigma_{yx}&=&g{\tilde e}^2 \beta\int \frac{d^3\vk}{(2\pi)^3}
	\tau_c\frac{\tau_c/\tau_B}{1+(\tau_c/\tau_B)^2}\frac{k_y^2}{\om^2} f_0(1\mp f_0)~,
	\nn\\
	&=&g{\tilde e}^2 \beta\int \frac{d^3\vk}{(2\pi)^3}
	\tau_c\frac{\tau_c/\tau_B}{1+(\tau_c/\tau_B)^2}\frac{\vk^2}{3\om^2} f_0(1\mp f_0)~,
	\nn\\
	\label{sig_xyCl}
	\eea
	where $g{\tilde e}^2=2\times 2\times 3\Big(\frac{4e^2}{9}+\frac{e^2}{9}+\frac{e^2}{9}\Big)=8e^2$ for 3 flavor quark matter~\cite{Dey:2019vkn,Ghosh:2019ubc}. For other relevant system or matter (e.g. HRG matter), we have to consider corresponding $g{\tilde e}^2$ values. 
	
	Similarly, $\sigma_{yy}$, $\sigma_{xy}$ can be obtained by repeating same calculation 
	for ${\vec E}=E_y{\hat y}$ and they are related as $\sigma_{xx}=\sigma_{yy}$, $\sigma_{xy}=-\sigma_{yx}$.
	Longitudinal conductivity along z-axis will remain unaffected by magnetic field, because
	Lorentz force does not work along the direction of the magnetic field.
	Hence, the classical expression of the longitudinal conductivity will be
	\bea
	\sigma_{zz} &=& g{\tilde e}^2 \beta\int \frac{d^3\vk}{(2\pi)^3}
	\tau_c\frac{k_z^2}{\om^2} f_0(1\mp f_0)~,
	\nn\\
	&=& g{\tilde e}^2 \beta\int \frac{d^3\vk}{(2\pi)^3}
	\tau_c\frac{\vk^2}{3\om^2} f_0(1\mp f_0)~,
	\label{sig_zzCl}
	\eea
	For particle and anti-particle cases, the term $\tau_c/\tau_B$ in the numerator for the Hall component, $\sigma_\times=\sigma_{xy}=-\sigma_{yx}$, has the opposite sign, resulting in vanishing net Hall conductivity at the vanishing quark/baryon chemical potential ($\mu=0$). Here, however, this fact may be difficult to see when looking at the Eq.~(\ref{sig_xyCl}), because the particle-anti-particle contribution is included in the degeneracy factor $g$ rather than the sum of their contributions. We will concentrate on the parallel component $\sigma_\parallel=\sigma_{xx}=\sigma_{yy}$ and the perpendicular component $\sigma_\perp=\sigma_{zz}$ because our current focus is on conductivity components of RHIC or LHC matter with almost zero quark/baryon chemical potential, where the Hall component vanishes. 
	
	\subsection{Quantum expressions of electrical conductivity}
	\label{sec:QM}
	Here, we will consider the effect of Landau quantization on the conductivity and see how it differs from CL expressions.
	The main modification will occur in the dispersion relation and the phase space integration. As we considered the magnetic field in the $z$ direction, momentum quantization will occur in the perpendicular plane, i.e. $x-y$ plane, so that,
	\begin{widetext}
		\bea
		\om = (\vk^2+m^2)^{1/2} &~~~\rightarrow~~~&  \om_{l} = (k_z^2+m^2+2l|{\tilde e}|B)^{1/2} ,
		\\
		2\int \frac{d^3\vk}{(2\pi)^3} &~~~\rightarrow~~~&  
		\sum_{l=0}^\infty \alpha_l \frac{|{\tilde e}|B}{2\pi} 
		\int\limits^{+\infty}_{-\infty} \frac{dk_z}{2\pi} ,
		\label{CM_QM}
		\eea
		where spin degeneracy 2 in left hand side of last line will be converted to $\alpha_l$,
		which will be 2 for all Landau levels $l$, except lowest Landau level (LLL) $l=0$, where
		$\alpha_l=1$. In general, one can write $\alpha_l = 2 - \delta_{l,0}$. Here, we also
		assume roughly, $k_x^2\approx k_y^2\approx (\frac{k_x^2+k_y^2}{2})=\frac{2l{\tilde e}B}{2}$,
		then conductivities can be expressed as
		\bea
		\sigma^{xx} = g {\tilde e}^2 \beta  \sum_{l=0}^\infty \alpha_l \frac{|{\tilde e}|B}{2\pi} 
		\int\limits^{+\infty}_{-\infty} \frac{dk_z}{2\pi} \frac{{l|{\tilde e}|B}}{\om^2_{l}} \tau_c 
		\frac{1}{1+(\tau_c/\tau_B)^2}
		f_0(\om_l)[1-f_0(\om_l)]
		\nn\\
		\sigma^{zz} = g {\tilde e}^2 \beta  \sum_{l=0}^\infty \alpha_l \frac{|{\tilde e}|B}{2\pi} 
		\int\limits^{+\infty}_{-\infty} \frac{dk_z}{2\pi} \frac{k^2_z}{\om^2_{l}} \tau_c 
		f_0(\om_l)[1-f_0(\om_l)]~.
		\label{Lsig_QM}
		\eea
	\end{widetext}
	Most previous works on Landau quantization~\cite{Hattori:2016cnt,Kurian:2018qwb,Kurian:2017yxj,Rath:2019vvi}, considered only longitudinal conductivity $\sigma^{zz}$ for $l=0$, known as LLL approximation. At extremely high magnetic fields, this scenario could be realised, in which all medium constituents occupy the lowest energy level $l=0$.  
	It means that perpendicular motion of medium constituents completely
	disappear as $k_x\approx k_y\approx 0$ at $l=0$. Therefore, in this LLL case, 
	$\sigma^{xx}\approx\sigma^{xy}\approx 0$ and $\sigma^{zz}\neq 0)$. However, below that strong
	magnetic field, $l>0$ energy levels might have some non-negligeable contributions, and the LLL approximation is not sufficient enough.
	%

	\subsection{Classical and quantum expression of electrical conductivity under the hardon resonance gas model}
	\label{sec:HRG}
	Now we will look at the HRG model calculation and see the transition from classical to quantum.
	The massless case is a non-interacting or Stefan-Boltzmann (SB) type limit type case, whereas HRG calculation maps the interacting picture. Electrical conductivity is not affected by neutral hadrons. So, for the conductivity calculation, we can
	classify hadrons into   
	\begin{enumerate}
		\item charged mesons (${M}$), which are basically bososns,
		\item charged baryons (${B}$), which are basically fermions.
	\end{enumerate}
	The contributions from both the types of hadrons will be added up. 
	Under the HRG model, the perpendicular and parallel components of electrical conductivity (Eqs.~(\ref{sig_xxCl}) and (\ref{sig_zzCl})) can be expressed as follows in the classical case:
	\bea
	\sigma^{xx} &=& \sum_{M,B}g{\tilde e}^2 \beta\int \frac{d^3\vk}{(2\pi)^3}
	\tau_c\frac{1}{1+(\tau_c/\tau_B)^2}\frac{\vk^2}{3\om^2} f_0(1\pm f_0)~,
	\nn\\
	\label{sxx_HRG_Cl}
	\\
	\sigma^{zz} &=& \sum_{M,B}g{\tilde e}^2 \beta\int \frac{d^3\vk}{(2\pi)^3}
	\tau_c\frac{\vk^2}{3\om^2} f_0(1\pm f_0)~.
	\label{szz_HRG_Cl}
	\eea

	Similarly, the quantum expressions Eqs.~(\ref{Lsig_QM}) will be modified as
	\begin{widetext}
		\bea
		\sigma^{xx} &=& \sum_{M}g {\tilde e}^2 \beta \Big(\frac{|{\tilde e}|B}{2\pi} \Big)  \sum_{l=0}^\infty\alpha_l  
		\int\limits^{+\infty}_{-\infty} \frac{dk_z}{2\pi} \frac{{(l+1/2)|{\tilde e}|B}}{\om^2_{l}} \tau_c 
		\frac{1}{1+(\tau_c/\tau_B)^2}
		f_0(\om_l)[1+f_0(\om_l)]
		\nn\\
		&+&\sum_{B}g {\tilde e}^2 \beta\Big(\frac{|{\tilde e}|B}{2\pi} \Big) \sum_{l=0}^\infty \alpha_l 
		\int\limits^{+\infty}_{-\infty} \frac{dk_z}{2\pi} \frac{{l|{\tilde e}|B}}{\om^2_{l}} \tau_c 
		\frac{1}{1+(\tau_c/\tau_B)^2}
		f_0(\om_l)[1-f_0(\om_l)]
		\label{sxx_HRG}
		\\
		\sigma^{zz} &=& \sum_{M}g {\tilde e}^2 \beta\Big(\frac{|{\tilde e}|B}{2\pi} \Big) \sum_{l=0}^\infty\alpha_l
		\int\limits^{+\infty}_{-\infty} \frac{dk_z}{2\pi} \frac{k^2_z}{\om^2_{l}} \tau_c 
		f_0(\om_l)[1+f_0(\om_l)]
		\nn\\
		&+&\sum_{B}g {\tilde e}^2 \beta\Big(\frac{|{\tilde e}|B}{2\pi} \Big) \sum_{l=0}^\infty \alpha_l 
		\int\limits^{+\infty}_{-\infty} \frac{dk_z}{2\pi} \frac{k^2_z}{\om^2_{l}} \tau_c 
		f_0(\om_l)[1-f_0(\om_l)]~.
		\label{szz_HRG}
		\eea
		%
		We have tabulated the expression of $\om_{l}$ and $\alpha_l$ for various particle species with their corresponding spin in the Table~(I)~\cite{Endrodi:2013cs,dePaoli:2012cz}. For the sake of simplicity, we only considered mesons with spin 0 and 1, as well as baryons with spin 1/2 and 3/2.
		Higher spin hadrons are not considered, which may be justified given their low thermodynamic weight factors due to their large masses. 
		\begin{table}[ht]
			\begin{center}
				\begin{tabular} {|c|c|c|c|}\hline
					$\mathrm{Particle \: \: species}$ & $\mathrm{Spin}$ & $\om_{l}$  & $\alpha_l$ \\
					\hline
					$\mathrm{Baryon}$  & $1/2$  & $\om_{l} = (k_z^2+m^2+2l|{\tilde e}|B)^{1/2}$   & $2-\delta_{l0}$  \\
					$\mathrm{Baryon}$  & $3/2$  & $\om_{l} = (k_z^2+m^2+2l|{\tilde e}|B)^{1/2}$   & $4-2\delta_{l0}-\delta_{l1}$  \\
					$\mathrm{Meson}$  & $0$  & $ \om_{l} = (k_z^2+m^2+(2l+1)|{\tilde e}|B)^{1/2}$   & $1$  \\
					$\mathrm{Meson}$  & $1$  & $ \om_{l} = (k_z^2+m^2+(2l+1)|{\tilde e}|B)^{1/2}$   & $3-\delta_{l0}$  \\
					\hline
					
				\end{tabular}
				\caption{Particles energy and degeneracy}
				\label{table1}
			\end{center}	
		\end{table}
		
	\end{widetext}
	Because baryons are fermions, their perpendicular momenta are $k_x^2\approx k_y^2\approx (\frac{k_x^2+k_y^2}{2})=\frac{2l{\tilde e}B}{2}$, whereas for the mesons perpendicular momenta are $k_x^2\approx k_y^2\approx (\frac{k_x^2+k_y^2}{2})=\frac{(2l+1){\tilde e}B}{2}$.
	As a result, there will be an interesting difference between massless quark matter and HRG systems.
	The perpendicular component will vanish for quark matter but not for HRG systems beacuse of non-zero contribution from the charged mesons in the LLL case.
	This fact will be illustrated graphically in the result section.    
	
	Here, for HRG system, we have to take care of the factors $g{\tilde e}^2$ for different hadrons,
	which are basically multiplication of degeneracy factor $g$ and their square of electric charge ${\tilde e}^2$.
	For example, $\pi^+$ meson with mass $140$ MeV, spin 0 has $g{\tilde e}^2=1\times e^2$; 
	$\rho^+$ meson with mass $770$ MeV, spin 1 has $g{\tilde e}^2=3\times e^2$;
	$\Delta^{++}$ baryon with mass $1232$ MeV, spin $3/2$ has $g{\tilde e}^2=4\times 4e^2$ etc.
	%

	\section{Results}
	\label{sec:Res}
	Let us begin with massless quark matter, which can be thought of as non-interacting or Stefan-Boltzmann (SB) limit type estimations. 
	We will not consider the gluons because they do not participate in conductivity.
	Massless quark matter entails taking into account all degeneracy factors such as three colour, two spin, two particle-anti-particle, and two flavour (with appropriate electric charge). 
	Though massless s quark consideration, like massless u and d quarks, is a bit of a rough approximation, QGP can reach these massless or SB limits at very high temperature limits. Following the estimation of the massless quark matter case, we will proceed to HRG calculations, which may be considered interacting QCD estimations because their thermodynamics match with LQCD thermodynamics data within the hadronic temperature range. Similar to thermodynamic quantities such as pressure, energy density, and entropy density, where massless or non-interacting or SB limit estimations are used as upper reference values, normalised conductivity components of massless quark matter will be used as that reference point here as well. Then, when we go for HRG estimates of conductivity components, we will be representing their interacting estimates, which are expected to be suppressed from their massless limits within the hadronic temperature domain, just like thermodynamical quantities. Estimation of these non-interacting to interacting  transformations of conductivity tensors are well discussed in Ref.~\cite{Dash:2020vxk} using the classical expressions of massless quark matter and HRG models, given in Eqs.~(\ref{sig_xxCl}), (\ref{sig_zzCl}), and (\ref{sxx_HRG_Cl}), respectively. 
	However, their transition from a classical to a quantum picture has never been explored in previous works using the quantum expressions given in Eqs.~(\ref{sxx_HRG}), (\ref{szz_HRG}). The purpose of this work is to investigate this fact. 
	So, to begin, we will investigate the classical to quantum transition for massless quark matter first, followed by HRG matter.
	Throughout the discussion, we will attempt to visualise all transitions—from massless or non-interacting pictures to HRG or interacting pictures, as well as from classical (CL) to quantum (QM) pictures. 
	\begin{figure} 
		\centering
		\includegraphics[scale=0.33]{sig_TM0.eps}
		\caption{Components of electrical conductivity for massless quark matter as a function of temperature.}
		\label{fig:el_T_m0}
	\end{figure}
	\begin{figure} 
		\centering
		\includegraphics[scale=0.33]{sig_BM0.eps}
		\caption{Components of electrical conductivity as a function of magnetic field for massless quark matter.}
		\label{fig:el_eB_m0}
	\end{figure}

	Let us begin by looking at Figs.~(\ref{fig:el_T_m0}) and (\ref{fig:el_eB_m0}), which show CL and QM curves of conductivity components along the $T$ and $B$ axes, respectively. Using CL Eq.~(\ref{sig_zzCl}), parallel component of conductivity for massless quark matter will be~\cite{Dey:2019axu} 
	\be 
	\sigma_{\parallel}=8e^2\frac{\zeta(2)}{3\pi^2}\tau_c T^2~,
	\ee 
	where $\zeta(2)=\pi^2/6$. 
	As a result, for the massless quark matter case, $\sigma_\parallel\propto \tau_c T^2$, so its normalized values $\sigma_\parallel/(\tau_c T^2)$ will be constant as shown by the green dash-dotted horizontal line in Figs.~(\ref{fig:el_T_m0}) and (\ref{fig:el_eB_m0}), which will serve as our reference points such as SB limit for thermodynamical quantities~\cite{Dash:2020vxk,Bali:2014kia}. 
	Next, we will calculate the deviation from the horizontal line in the high $B$ and low $T$ domains using QM Eq.~(\ref{Lsig_QM}) of $\sigma_\parallel$. 
	This fact is explored by the QM curves of $\sigma_\parallel$ - grey double-dash-dotted lines in Figs.~(\ref{fig:el_T_m0}) and (\ref{fig:el_eB_m0}). 
	This deviation increases towards the low $T$ and high $eB$ domains due to Landau quantization, the transition from integration to Landau level summation, as denoted in Eq.~(\ref{CM_QM}).
	The QM and CL curves of $\sigma_\parallel$, on the other hand, are merging due to integration and Landau level summation equivalence in the high $T$ and low $eB$ ranges. 
	In the left panel of Fig. (10) in Ref.~\cite{Bali:2014kia}, a similar enhancement for entropy density with respect to SB limit was found in the low $T$ domain. 
	Therefore, the SB or massless limits of normalised entropy density ($s/T^3$) or other thermodynamical quantities, as well as normalised longitudinal conductivity ($(\sigma_\parallel/\tau_c T^2)$, exhibit the same pattern--their horizontal lines are deviated (enhanced) in the low $T$ domain due to Landau quantization. 
	This Landau quantization effect is can be seen in low $T$ and high $eB$ domains, which is a well-known fact for all thermodynamical quantities and transport coefficients.
	Based on this, we can assume that low $T$ and high $eB$ are quantum domains whereas high $T$ and low $eB$ are classical domains.
	Using these normalised quantities ($s/T^3$, $\sigma_\parallel/\tau_c T^2$) of massless quark matter, we observe an enhancement of magneto-thermodynamical phase-space in the low $T$ and high $eB$ domains.  
	
	We also estimated $\sigma_\parallel/\tau_c T^2$ for the lowest Landau level (LLL) approximation by imposing $l=0$ in Eq. (\ref{Lsig_QM}), as shown in Figs. (\ref{fig:el_T_m0}) and (\ref{fig:el_eB_m0}) by cyan dash-double dotted. 
	The enormous difference between the full Landau level estimation (brown double-dash-dotted lines) and the LLL approximation (cyan dash-double-dotted) is visible. As expected, they are merging into the high $eB$ and low $T$ domains. 
	However, according to our graphs, this converence of LLL approximation and actual QM estimation is unlikely within the range of RHIC/LHC matter, say $T = 0.1-0.4$~GeV and $eB = 1-10 m_\pi^2$.   
	From Fig.~(\ref{fig:el_eB_m0}), we can say that for massless QGP at $T=0.150$ GeV, the Landau quantization effect is noticeable beyond $eB=1~m_\pi^2$ and LLL is not a good approximation within $eB=(1-10)m_\pi^2$.  As a result, neither the CL nor the LLL approximation will be valid for estimating conductivity for RHIC/LHC matter with an expected magnetic field of $eB = 1-10~m_\pi^2$.
	The current work is intended to convey the message. 
	
	The perpendicular conductivity $\sigma_\perp$, from Eq. (\ref{sig_xxCl}) can be simplified for massless quark matter as 
	\be 
	\sigma_{\perp}=12\frac{\zeta(2)}{3\pi^2}\tau_c T^2\sum_{f=u,d,s}\frac{e_f^2}{1+(\tau_ce_fB/3T)^2}~,
	\ee 
	where $e_f=+\frac{2}{3}e$, $-\frac{1}{3}e$, $-\frac{1}{3}e$ for $f=u, d, s$ and average energy of massless quark is $3T$. 
	To grasp the graphical nature of $\sigma_\perp$, we looked at the simple analytic form, which has an additional anisotropic factor $\frac{1}{1+(\tau_c/\tau_B)^2}$ with an average of $\tau_B=3T/(e_fB)$. 
	As a result, the dimensionless value of $\sigma_\perp/(\tau_cT^2)$ is an anaisotropic factor that rises with $T$ and falls with $eB$. The solid black lines in Figs.~(\ref{fig:el_T_m0}) and (\ref{fig:el_eB_m0}) demonstrate this fact.  In comparison to the CL curves, the QM curves of $\sigma_\perp/(\tau_cT^2)$(red dash line) are suppressed, whereas the QM curves of $\sigma_\parallel/(\tau_cT^2)$ (brown double-dash-dotted line) are enhanced.
	Landau quantization of the FD distribution, in fact, always enhances probability in the quantum domain (low $T$ and high $eB$), resulting in higher $\sigma_\parallel/(\tau_c T^2)$.
	While Landau quantization increases $\sigma_\perp/(\tau_cT^2)$ via FD distribution, the anisotropic factor reduces it.
	Their combined effect yields unprecedented values in the quantum domain.

	\begin{figure} 
		\centering
		\includegraphics[scale=0.33]{sig_T.eps}
		\caption{Components of electrical conductivity for HRG matter as a function of temperature.}
		\label{fig:el_T_hrg}
	\end{figure}
	\begin{figure} 
		\centering
		\includegraphics[scale=0.68]{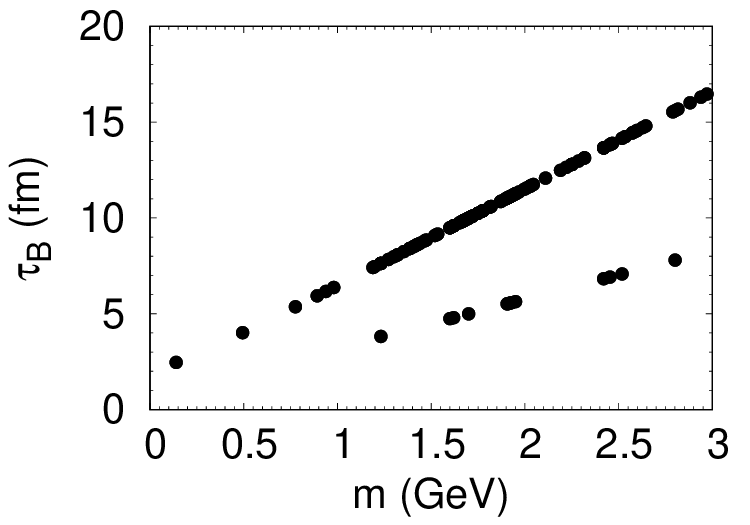}
		\caption{Average values of inverse cyclotron frequency $\tau_B$ (in fm) for different hadrons.}
		\label{fig:tB_m_hrg}
	\end{figure}
	\begin{figure} 
		\centering
		\includegraphics[scale=0.33]{sig_B.eps}
		\caption{Components of electrical conductivity as a function of magnetic field in HRG model}
		\label{fig:el_eB_hrg}
	\end{figure}
	%
	%
	
	Let us now proceed to the HRG model conductivity estimation, where all charged hadrons will be included with their respective statistical weight factors due to their different mass values.  
	These different statistical weight factors (roughly $e^{-\beta m}$) of the FD and BE distributions will act on baryons and mesons, causing HRG estimations to remain lower than massless quark matter. 
	According to Fig.~(\ref{fig:el_T_hrg}), in the hadronic temperature domain $T =0.100-0.170$ GeV, $\sigma_{\parallel,\perp}/(\tau_c T^2)$ for HRG estimation, the range is 0.003-0.01, which is significantly lower than their masless quark matter estimation (around $0.02$).  
	Similar suppression values in thermodynamical quantities are realised as non-perturbative sources of QCD in LQCD calculations, which is well mapped alternatively by the HRG model. 
	Therefore, we can assume that Figs.~(\ref{fig:el_T_m0}) and (\ref{fig:el_T_hrg}) reflect the pQCD/massless limits and non-pQCD domains of conductivity, respectively, in the high and low temperature domains. 
	This fact is thoroughly examined in Ref.~\cite{Dash:2020vxk} using CL expressions of the massless and HRG systems. 
	Here, we should concentrate only on its QM modification for the purposes of this work, and some interesting results are as follows. 
	QM $>$CL in the low $T$ and high $eB$ domains for the massless case, but QM $<$ CL for the HRG case for $\sigma_\parallel/(\tau_c T^2)$. 
	To understand the suppressed QM values for HRG estimation, where both bosons and mesons contribute, their individual investigations will be helpful. That will be discussed later.   
	
	To understand the detailed cyclotron motions of different charged hadrons, we have calculated their average cyclotron time period (inverse of cyclotron frequency)
	\be
	\tau_B (m) = \frac{\om_{\rm av}}{|{\tilde e}|B} =\frac{1}{|{\tilde e}|B}\frac{\int d^3\vk ~\om f_0(m)}{\int d^3\vk ~f_0(m)},
	\ee 
	which is plotted against the mass $(m)$ axis in Fig.~\ref{fig:tB_m_hrg}. Different points represent different hadrons in their respective masses. 
	It is obvious that $\tau_B$ will increase with $m$ and decrease with $\tilde e$, as shown by the graphs.
	The majority hadrons with charges $|\tilde e|=|e|$ lie within a specific slope of $\tau_B$ vs $m$, while a few deviate due to their charges $|\tilde e|>|e|$. 
	As a result, higher charged, lighter mass hadrons will play an important role in the suppression of  $\sigma_\perp$ via the anisotropic factor $\frac{1}{1+(\tau_c/\tau_B)^2}$.

	We plotted $\sigma_\perp$ and $\sigma_\parallel$ as a function of magnetic field in Fig.~\ref{fig:el_eB_hrg} at $T=0.150$~GeV and $\tau_c=1$~fm  for Cl, QM, and the LLL cases in the HRG system.   
	%
	\begin{figure}
		\centering
		\includegraphics[scale=0.33]{P-pi.eps}
		\caption{Perpendicular component of electrical conductivity as a function of magnetic field for pion and proton}
		\label{fig:el_eB_had}
	\end{figure}
	The Table~(\ref{table2}) summarises the key findings by comparing the HRG estimations in Figs.~(\ref{fig:el_T_hrg}) and (\ref{fig:el_eB_hrg}) with massless quark matter estimations in  Figs.~(\ref{fig:el_T_m0}) and (\ref{fig:el_eB_m0}). This applies to the hadronic temperature range $T\approx0.100$-$0.170$ GeV with experimentally expected magnetic field range $eB\approx(0$-$10)m_\pi^2$,
	and more effective in the quantum domain, i.e., low $T$ and high $eB$ zone.
	When we individually analyze proton and pion matter contributions, we can understand why massless quark matter and HRG matter outcomes are quite different from one another.  
	These are discussed in the next paragraphs for the case of $\sigma_{\parallel}$ only.
	
	\begin{table}[ht]
		\begin{center}
			\begin{tabular} {|c|c|c|}\hline
				Medium & $\sigma_{\parallel}(T,B)$ & $\sigma_{\perp}(T,B)$ \\
				\hline
				Massless quark matter & QM $>$ CL &  QM $<$ CL \\
				HRG matter & QM $<$ CL &  QM $>$ CL \\	
				Proton matter & QM $>$ CL &  QM $<$ CL \\
				Pion matter & QM $<$ CL &  QM $>$ CL \\
				\hline
			\end{tabular}
			\caption{Comparison between QM and CL estimations for different cases.}
			\label{table2}
		\end{center}	
	\end{table}
	
	The tabulated outcome indicates that the quantum effects on pion ($\pi$) and pron ($P$) are exactly opposite due to their corresponding bosonic and fermionic distribution functions, respectively. 
	Moreover, the qualitative results of HRG and pion matter are identical, indicating that mesons dominate baryons in the estimation of HRG. 
	Mesons are expected to dominate in the HRG estimations due to their higher thermodynamical probabilities as a result of their low masses.
	Here, we explore this fact graphically for one of the components of conductivity, $\sigma_\perp$. 
	In Fig.~\ref{fig:el_eB_had}, we have plotted $\sigma_{\perp}/(\tau_c T^2)$ as a function of $eB/m_\pi^2$ for a pion (upper panel) and a proton (lower panel) at $T=0.150$~GeV and $\tau_c=1$~fm. 
	We find that the conductivity for $\pi^+$ is much higher than that of p, which is because of the small mass of pion. 
	Although the QM results for the pion case (red dashed line) are larger than the CL results (black solid line), the difference is negligible.
	In contrast, the opposite ranking is clearly observed in the proton case. 
	The magneto-thermodynamical phase spaces of the pion and proton, which are bosons and fermions, are enhanced and suppressed, respectively, due to their (Landau) quantized BE and FD distribution functions. 

	\begin{figure} 
		\centering
		\includegraphics[scale=0.3]{QM_CL.eps}
		\caption{Percentage deviation of classical estimation of conductivity from that of quantum, as a function of magnetic field. Deviation $=\frac{\sigma^{\rm QM}_\perp-\sigma^{\rm Cl}_\perp}{\sigma^{\rm QM}_\perp}$}
		\label{fig:Cl-QM}
	\end{figure}
	%
	%
	The Fig.~\ref{fig:Cl-QM} provides a detailed description of how pion and proton matter is either enhanced or suppressed in QM results compared to CL results.  
	Here, we have plotted $\frac{\sigma^{\rm QM}_\perp-\sigma^{\rm Cl}_\perp}{\sigma^{\rm QM}_\perp}\times 100\%$ as a function of $eB/m_\pi^2$ for HRG (black solid line), $\pi^+$ (red dotted line), and p (blue dash line) matters. 
	Notice that below $eB=1m^2_\pi$, the deviation is less than $1\%$. 
	For the HRG result, deviation increases slowly with the magnetic field, reaches a maximum of $\sim 3\%$ and then decreases to $0\%$ at $eB\approx 10 m_\pi^2$, and further increase of magnetic field will start the suppression. 
	Even though the CL result for pion and proton is deviated from the QM result, they follow a similar trend, and their combined effect is reflected in the HRG result. 
	Within the magnetic field range $eB=0$-$10 m_\pi^2$ and temperature $T=0.150$ GeV, the contribution from mesons outweighs that from baryons. 
	As a result, the HRG results exhibit effectively bosonic quantum modification, which is why HRG results are the opposite to those of massless quark matter, which are subject to fermionic quantum modification.

	
	\section{Summary} 
	\label{sec:Sum}
	In summary, we have explored comparative estimations of classical and quantum expressions of electrical conductivity in the presence of a magnetic field. 
	We first obtained the results for the massless quark gluon plasma, and then we used HRG model calculations to obtain the results for interacting QCD. 
	In the presence of a magnetic field, there will be three components: parallel, perpendicular, and Hall.
	At zero quark/baryon chemical potential, the medium carries an equal number of opposite electrical charges, so the Hall conditions will disappear.
	In both classical (without considering Landau quantization) and quantum pictures, parallel and perpendicular conductivity become different in the presence of the external magnetic field.
	Conductivities change from being isotropic to anisotropic with a low to high magnetic field in both the CL and QM cases.  
	In Refs.~\cite{Dash:2020vxk}, the classical estimation of conductivity tensors for massless quark matter and HRG matter is thoroughly discussed.  
	The goal of the current work is to discover their quantum extension by introducing Landau quantization. 
	Our estimation is limited to the hadronic temperature domain $T=0.100-0.170$ GeV and the magnetic field domain $eB=0-10 m_\pi^2$. The findings are summarised in the bullet points below.
	\begin{itemize}
		\item QM enhancement is observed in parallel/longitudinal conductivity for massless quark matter and proton (baryonic) matter in the presence of a magnetic field.
		\item QM suppression is observed in perpendicular/transverse conductivity for massless quark matter and proton (baryonic) matter in the presence of a magnetic field.
		\item QM suppression is observed in parallel/longitudinal conductivity for pionic (mesonic) matter and HRG matter in the presence of a magnetic field. 
		\item QM enhancement is observed in perpendicular/transverse conductivity for pionic (mesonic) matter and HRG matter in the presence of a magnetic field.
	\end{itemize}
	The QM effects in bosons and fermions are different because of their corresponding BE and FD distribution functions, which also serve to reveal the effect of Landau level summation in two different ways.
	Additionally, we have demonstrated that, for the ranges of temperature and magnetic field under consideration, the lowest Landau level approximation of the HRG estimation is in no way a good approximation.  
	In the presence of a magnetic field, full Landau level summation is recommended for the quantum version of HRG estimations.
	Therefore, the current work, for the first time, suggests a non-negligible quantum effect in the estimation of the transport coefficient for HRG matter, which may also be applicable to other HRG-based phenomenology.
	In Refs.~\cite{Kurian:2018qwb,Kurian:2020qjr}, where a quasiparticle-based model (effective fugacity quasiparticle model) was used to estimate the transport coefficients, a similar effect of Landau quatization for QGP medium was also found.
	The current work has only considered the electrical conductivity, but we anticipate it will be applicable to other transport coefficients such as shear and bulk viscosity, thermal conductivity, as well as other HRG-related phenomenology that may be investigated in the future.

	{\bf Acknowledgment:} 
	S. Satapathy acknowledges to the fellowship, funded by DST INSPIRE Faculty scheme of Research Project (IFA18-PH220) and to principle investigator of that project (Dr. Sudipan De).

\end{document}